\documentclass[prd,preprint,floatfix,nofootinbib,amsmath]{revtex4}
\usepackage{epsfig}
\usepackage{graphicx}
\usepackage{pstricks}

\catcode`\@=11
\def\sla@#1#2#3#4#5{{%
 \setbox\z@\hbox{$\m@th#4#5$}%
 \setbox\tw@\hbox{$\m@th#4#1$}%
 \dimen4\wd\ifdim\wd\z@<\wd\tw@\tw@\else\z@\fi
 \dimen@\ht\tw@
 \advance\dimen@-\dp\tw@ \advance\dimen@-\ht\z@
 \advance\dimen@\dp\z@
 \divide\dimen@\tw@ \advance\dimen@-#3\ht\tw@
 \advance\dimen@-#3\dp\tw@ \dimen@ii#2\wd\z@
 \raise-\dimen@\hbox to\dimen4{%
 \hss\kern\dimen@ii\box\tw@\kern-\dimen@ii\hss}%
 \llap{\hbox to\dimen4{\hss\box\z@\hss}}}}

\def\slashed#1{%
 \expandafter\ifx\csname sla@\string#1\endcsname\relax
{\mathpalette{\sla@/00}{#1}}
\fi}
\def\declareslashed#1#2#3#4#5{%
 \expandafter\def\csname sla@\string#5\endcsname{%
#1{\mathpalette{\sla@{#2}{#3}{#4}}{#5}}}}
 \catcode`\@=12
\declareslashed{}{/}{.08}{0}{D}
 \declareslashed{}{/}{.1}{0}{A}
 \declareslashed{}{/}{0}{-.05}{k}
 \declareslashed{}{/}{.1}{0}{\partial}
 \declareslashed{}{\not}{-.6}{0}{f}
\def\lsim{\mathrel {\vcenter {\baselineskip 0pt \kern 0pt
    \hbox{$<$} \kern 0pt \hbox{$\sim$} }}}
\def\gsim{\mathrel {\vcenter {\baselineskip 0pt \kern 0pt
    \hbox{$>$} \kern 0pt \hbox{$\sim$} }}}
\def\slashchar#1{\setbox0=\hbox{$#1$}           % set a box for #1
 \dimen0=\wd0                                 % and get its size
  \setbox1=\hbox{/} \dimen1=\wd1               % get size of /
\ifdim\dimen0>\dimen1                        % #1 is bigger
  \rlap{\hbox to \dimen0{\hfil/\hfil}}      % so center / in box
  #1                                        % and print #1
  \else                                        % / is bigger
 \rlap{\hbox to \dimen1{\hfil$#1$\hfil}}   % so center #1
   /                                         % and print /
  \fi}                                         %
\def\cpto{\mathrel {\vcenter {\baselineskip 0pt \kern 0pt
    \hbox{$CP$} \kern 0pt \hbox{$\longrightarrow$} }}}
\def\cptof{\mathrel {\vcenter {\baselineskip 0pt \kern 0pt
    \hbox{$~CP$} \kern 0pt \hbox{$\longleftrightarrow$} }}}

\begin{document}

\baselineskip=15pt
\preprint{}

\title{$CP$ violating anomalous top-quark couplings at the LHC}

\author{Sudhir Kumar Gupta, Alaettin Serhan Mete and G. Valencia}

\email{skgupta@iastate.edu,serhan@iastate.edu,valencia@iastate.edu}

\affiliation{Department of Physics, Iowa State University, Ames, IA 50011.}

\date{\today}

\vskip 1cm
\begin{abstract}

We study the $T$ odd correlations  induced by $CP$ violating anomalous top-quark couplings at both production and decay level in the process $gg \to t\bar{t} \to (b\mu^+ \nu_\mu) (\bar{b}\mu^- \bar{\nu}_\mu)$. We consider several counting asymmetries at the parton level and find the ones with the most sensitivity to each of these anomalous  couplings at the LHC. 

\end{abstract}


\maketitle

\section{Introduction}

With the upcoming start of the LHC in mind, we consider the possibility of looking for $CP$ violation in high energy processes involving the production and decay of top quarks. There are many observables suitable for this purpose that have been studied before in the literature \cite{reviews, Donoghue:1987ax, Ma:1991ry,Bernreuther:1992be,Brandenburg:1992be,Atwood:1992vj,Bernreuther:1993hq,Choi:1997ie,Zhou}. In this paper, we consider $T$ odd triple product correlations of the sort first discussed in Ref.~\cite{Donoghue:1987ax}. 

We keep our study as model independent as possible by considering a scenario in which only the standard model (SM) particles are relevant and the $CP$ violation is parametrized by anomalous top-quark couplings affecting both the production and decay  vertices \cite{Bernreuther:1992be}.  The $t\bar{t}$ production process is modified relative to the SM by the interaction
\begin{eqnarray}
{\cal L}_{cdm}&=&-ig_s\frac{\tilde{d}}{2}\bar{t}\, \sigma_{\mu\nu}\gamma_5  \, G^{\mu\nu}\, t,
\label{dtilde}
\end{eqnarray}
where $g_s$ is the strong coupling constant and $G^{\mu\nu}$ is the usual gluon field strength tensor. This interaction results in a modification to the $t\bar{t} g$ vertex as well as in an additional, ``seagull'', $t\bar{t} gg$ vertex.
For the decay process $t\to b W^+$ we write the most general decay vertex  (and  the corresponding one for  $\bar{t}$ decay), 
\begin{eqnarray}
\Gamma^\mu_{Wtb} &=& 
-\frac{g}{\sqrt{2}} \, V_{tb}^\star \,\bar{u}(p_b) \left[ \gamma_\mu (f_1^L P_L+f_1^R P_R)-
i  \sigma^{\mu\nu} (p_t-p_b)_\nu (f_2^L P_L+f_2^R P_R)\right] u(p_t). \label{ftilde}
\end{eqnarray}
This vertex can be derived from a dimension five effective Lagrangian as in Ref.~\cite{delAguila:2002nf}, but unlike the case of Eq.~\ref{dtilde}, the effective Lagrangian does not generate other vertices that affect this calculation. For the remainder of this paper we will use $V_{tb}\equiv 1$, $f_1^L=1$, $f_1^R=0$ and $f_2^L=0$ as in the SM, and allow for new physics only through the coupling $f_2^R$ which is the only one that can interfere with the SM to produce $T$-odd correlations.To generate  $T$-odd observables the coupling $f_2^R$ must have a phase but this phase does not have to be $CP$ violating. We thus write $f_2^R=f\exp{i(\phi_f+\delta_f)}$ using $\phi_f$ to parametrize a $CP$ violating phase due to new physics and $\delta_f$ a $CP$ conserving phase arising from real intermediate states at the loop level. Some of these $CP$ violating couplings have also been studied in the context of a future linear collider~\cite{eeanom}.
 
Our study in this paper corresponds to the numerical implementation of the results presented in Ref.~\cite{Antipin:2008zx}, where comparisons to the previous literature were made \cite{Brandenburg:1992be,Atwood:1992vj,Choi:1997ie}. 
We start from Eqs.~10~and~32 of Ref.~\cite{Antipin:2008zx} which give the spin and color averaged matrix element squared containing the $T$ odd triple product correlations for the process $gg \to t {\bar t} \to b \mu^+ \nu_\mu {\bar b} \mu^-\bar{\nu}_\mu$.  For convenience we collect the results of Ref.~\cite{Antipin:2008zx}  in the Appendix.
Numerical studies similar to ours have been carried out before for a subset of the observables we consider here, and we compare our results to the most recent ones in the literature \cite{Sjolin:2003ah, AguilarSaavedra:2007rs}. The numerical studies are performed with the aid of MADGRAPH \cite{Stelzer:1994ta,Alwall:2007st,Alwall:2008pm}. 

\section{Observables}

In Ref.~\cite{Antipin:2008zx} three independent correlations were identified to test $CP$ violation in $t\bar{t}$ production. Here we rewrite them with a small modification: we have reshuffled factors of $t-u$ between the form factors and the correlations to ensure that both the form factor and the correlation are even under the interchange of the two initial protons. We concentrate initially on $W$ decaying into muons, for which the $T$-odd correlations are\footnote{Here we use the Levi-Civita tensor contracted with four vectors $\epsilon(a,b,c,d) \equiv \epsilon_{\mu \nu \alpha \beta} a^\mu b^\nu c^\alpha d^\beta$ with the sign convention $\epsilon_{0123}=1$. We also use $s,t,u$ to refer to the parton level Mandelstam variables for $gg \to t \bar{t}$.}:
\begin{eqnarray}
{\cal O}_1 &=& \epsilon(p_t,p_{\bar{t}},p_{\mu^+},p_{\mu^-}) \nonumber \\
{\cal O}_2 &=& \,(t-u) \,\epsilon(p_{\mu^+},p_{\mu^-},P,q) \nonumber \\
{\cal O}_3 &=&\,(t-u) \,\left( P \cdot p_{\mu^+} \,
\epsilon(p_{\mu^-},p_t,p_{\bar{t}},q)+P \cdot p_{\mu^-} 
\,\epsilon(p_{\mu^+},p_t,p_{\bar{t}},q) \right)
\label{prodco}
\end{eqnarray}
with $q=p_1- p_2$ and $P=p_1+p_2$ being the difference and sum of the incoming parton momenta. The spin and color averaged matrix element squared that contains these correlations is given by,
\begin{eqnarray}
\left| {\cal M}\right|_{CP}^2  &=&  \, C_1(s,t,u) \, {\cal O}_1 \,+\, C_2(s,t,u) \, {\cal O}_2 \,+\, C_3(s,t,u)\,{\cal O}_3, 
\label{formfactors}
\end{eqnarray}
where the form factors $C_{1,2,3}$ were computed in Ref.~\cite{Antipin:2008zx} and we reproduce them in the appendix for convenience.  

We begin by studying each of the three terms in Eq.~\ref{formfactors} separately, considering  the lab frame distributions $d\sigma/d{\cal O}_i$ for the three correlations. 
In each case we  isolate the $CP$ odd form factor $C_i$ by constructing the integrated counting asymmetry 
\begin{eqnarray}
A_i &\equiv & \frac{N_{events}({\cal O}_i >0)-N_{events}({\cal O}_i<0)}{N_{events}({\cal O}_i>0)+N_{events}({\cal O}_i<0)}.
\label{asym}
\end{eqnarray}

The observables used to construct the $A_i$ are not realistic in that not all the momenta appearing in them can be reconstructed. To  address this issue we replace those observables assuming that for each event it is only possible to reconstruct the momenta of the two muons $\mu^\pm$, the two $b, \bar{b}$ jets, and the beam direction. The correlations under this assumption can be obtained from Eq.~\ref{prodco} with the substitutions
\begin{eqnarray}
p_t \to  p_b + p_{\mu^+} && p_{\bar t} \to p_{\bar b} + p_{\mu^-} \nonumber \\
P \to   p_b + p_{\mu^+} + p_{\bar b} + p_{\mu^-} &&
q \to  \tilde{q} \equiv P_1-P_2.
\label{repls}
\end{eqnarray}
We have defined a four-vector $\tilde{q}$, as the difference between the two beam four-momenta.  The factor $t-u$ could get modified by writing it as $(t-u) = q\cdot (p_{\bar t}-p_t)$ with the substitutions implied by Eq.~\ref{repls}. However,  all one needs is a factor linear in $\tilde{q}$ so we choose the simpler form $(t-u) \to  \tilde{q}\cdot (p_{\mu^-}-p_{\mu^+})$.

All this results in the correlations $\tilde{\cal{O}}$, 
\begin{eqnarray}
\tilde {\cal{O}}_1 &=& \epsilon(p_b,p_{\bar{b}},p_{\mu^+},p_{\mu^-}) \nonumber \\
\tilde {\cal{O}}_2 &=& \, \tilde{q}\cdot (p_{\mu^+}-p_{\mu^-}) \,\epsilon(p_{\mu^+},p_{\mu^-},p_b+p_{\bar{b}},\tilde{q}) \nonumber \\
\tilde {\cal{O}}_3 &=& \, \tilde{q}\cdot (p_{\mu^+}-p_{\mu^-}) \,\epsilon(p_{b},p_{\bar b},p_{\mu^+}+p_{\mu^-},\tilde{q}),
\label{prodcoprime}
\end{eqnarray}
and their associated  counting asymmetries  $\tilde{A}_i$. 
It is easy to see that  the correlation  ${\cal O}_3$ gives rise to both $\tilde {\cal O}_2$ and $\tilde {\cal O}_3$. From the experimental perspective, $\tilde {\cal{O}}_2 $ is most desirable as it is the only one that does not require distinguishing between the $b$ and $\bar {b}$ jets. 

When $CP$ violation occurs in the decay vertex, the spin and color averaged matrix element squared containing the $T$-odd correlations was written in Ref.~\cite{Antipin:2008zx} as~\footnote{Note that there is a typo in Ref.~\cite{Antipin:2008zx} where $\phi_f$ and $\delta_f$ are reversed.}
\begin{eqnarray}
|{\cal M}|^2_{T} &=& \, 
f\sin(\phi_f+\delta_f)\, \epsilon(p_t,p_{ b},p_{\ell^+},Q_{t}) +
 f\sin(\phi_f-\delta_f) \,\epsilon(p_{\bar t},p_{ \bar{b}},p_{\ell^-},Q_{\bar{t}}) .
 \label{asymcpdec}
\end{eqnarray} 
All the terms in Eq.~\ref{asymcpdec} contain three four-momenta from one of the decay vertices so the correlations ${\cal O}_{1,2,3}$ defined previously may not be the best to measure these couplings. Guided by the form of Eq.~\ref{asymcpdec}, we define the following three correlations for this purpose
\begin{eqnarray}
{\cal O}_4 &=& \epsilon(P,p_b-p_{\bar{b}},p_{\mu^+},p_{\mu^-}) \nonumber \\
{\cal O}_5 &=& \epsilon(p_t,p_{\bar t},p_b+p_{\bar{b}},p_{\mu^+}-p_{\mu^-}) \nonumber \\
{\cal O}_6 &=&\,(t-u) \,
\epsilon(P,p_b+p_{\bar{b}},p_{\mu^+}-p_{\mu^-},q).
\label{decayco}
\end{eqnarray}
These three correlations explicitly show that in order to test for $CP$ violation in the decay vertex it is necessary to compare the decay of the top quark with that of the anti-top quark. This is accomplished in Eq.~\ref{decayco} with the use of the linear combinations of $p_b\pm p_{\bar{b}}$ as well as $p_{\mu^+}- p_{\mu^-}$. These constructions are $CP$-odd, and as such they isolate the phase $\phi_f$ in Eq.~\ref{asymcpdec}. Later on we discuss alternative constructions to isolate the phase $\delta_f$. Interestingly, after we use 
the substitutions of Eq.~\ref{repls} to account for the fact that the top four-momenta cannot be reconstructed completely,  no new correlations are needed: both ${\cal{O}}_{4,5}$ become proportional to $\tilde{\cal{O}}_1$ and  ${\cal{O}}_6$  becomes proportional to $\tilde{\cal{O}}_2$.

\section{Numerical Analysis}

We start from the standard model process $gg\to t{\bar t} \to b \mu^+ \nu_\mu {\bar b} \mu^-\bar{\nu}_\mu$ implemented in MADGRAPH according to the decay chain feature described in Ref.~\cite{Alwall:2008pm}. This decay chain feature is chosen for consistency with the approximations in the analytical calculation of the $CP$ violating interference term presented in Ref.~\cite{Antipin:2008zx}, in which the narrow width approximation is used for the intermediate top quark and $W$ boson states. The expressions from Ref.~\cite{Antipin:2008zx} are then added to the spin and color averaged matrix element squared for the SM (which MADGRAPH calculates automatically) and the resulting code is used to generate events. This code is then missing the terms that are completely due to new physics: those proportional to the anomalous couplings squared. This approximation is justified because those terms do not generate $T$-odd correlations. In addition, as long as the conditions that allow us to write the new physics in terms of anomalous couplings remain valid, their contribution to the total cross-section is small.

For event generation we use the default MADGRAPH cuts requiring the top quark and $W$ boson intermediate states to be within 15 widths of their mass shell, the $p_T$ of both muons to be larger than 10~GeV and $\eta_\mu < 2.5$. We also use SM parameter values as in Madgraph, except for $m_b=0$; and we use the CTEQ-6L1 parton distribution functions. Furthermore, we only include the gluon fusion initiated parton processes. This procedure leads to a cross-section $\sigma(pp\to t\bar{t}\to b \mu^+ \nu_\mu \bar{b} \mu^- \bar{\nu}_\mu) \approx 4.3~{\rm pb}$, which is independent of the anomalous top-quark couplings $\tilde{d},~f$. The total cross-section does not depend on $\tilde{d},~f$ because we are only keeping deviations from the SM that are $T$ odd. These terms are also $P$ odd, and as such they integrate to zero.

The 4.3~pb cross-section we obtain underestimates the known theoretical cross-section by almost a factor of two. The latter is derived by multiplying the next-to leading order (NLO) total cross-section for $t\bar{t}$ production at the LHC with the $W\to \mu\nu$ branching ratios: $833~{\rm pb} \times (0.1057)^2 \approx 9.3~{\rm pb}$. About $20\%$ of this difference arises from the NLO corrections and most of the rest from our decay chain approximation. We expect the asymmetries we compute to change slightly when the decay chain approximation is relaxed and NLO corrections are included.  Numerically we have checked that the asymmetry increases with the cross-section as we relax the decay chain approximation. We cannot quantify rigorously this increase  because our starting (analytic) formula was computed in the narrow width approximation. In addition,  since the asymmetry is an interference effect between the new physics and the SM amplitude, we expect it will get modified by about $10\%$ by NLO QCD corrections. 

We first estimate the counting asymmetries of Eq.~\ref{asym}  by generating $10^6$ events for each of the four cases: $\tilde{d} = 5 \times 10^{-4}~{\rm GeV}^{-1}$; $f\sin\phi_f= 5 \times 10^{-4}~{\rm GeV}^{-1}$; $f\sin\delta_f = 5 \times 10^{-4}~{\rm GeV}^{-1}$ and $\tilde{d} = f =0$. These cases correspond to $CP$ violation in the production vertex, $CP$ violation in the decay vertex, strong phases in the decay vertex and the lowest order SM respectively. According to our discussion in Appendix~\ref{a:stat},  we require that the signals be larger than the $3\sigma$ level for statistical fluctuations, $A_i  \sim 3 \times 10^{-3}$.  Asymmetries found to be at this level or below, are estimated once again by increasing $\tilde{d}$ or $f$ by factors of ten,  to $5 \times 10^{-3}~{\rm GeV}^{-1}$. If the results scale linearly with $\tilde{d}$ (or $f$), they are treated as true $CP$-odd asymmetries and entered into Table~\ref{t:basic}. Otherwise they are treated as zero and examined further in Appendix~\ref{a:stat} by increasing the number of generated events and/or the size of the coupling. All the zero entries in Table~\ref{t:basic} satisfy our a-priori expectations. All the observables ${\cal{O}}_{1-6}$ are truly $CP$ odd, so they do not get contributions from the SM ($\tilde{d}, {f} =0$) nor from the final state interactions ($\sin\phi_f=0$ but $\sin\delta_f \neq 0$). The case  of $A_{1-3}$ as induced by ${f}\sin\phi_f$ is not clear cut as seen in  Appendix~\ref{a:stat}, but we do not pursue it further since there are larger, more promising, asymmetries to constrain this coupling.
%%%%%%%%%%%%
\begin{table}[h]
\centering
\begin{tabular}{| c | c | c | c | c | c | c |}
\hline\hline
 & $A_1$ & $A_2$ & $A_3$ & $A_4$ & $A_5$ & $A_6$\\
\hline 
$\tilde{d}$      & $9.4\times 10^{-2}$ & $-2.3 \times 10^{-2}$ & $2.0 \times 10^{-3}$ &$-6.9 \times 10^{-2}$ &$3.4\times 10^{-2}$ &$-8.1 \times 10^{-3}$ \\
${f}\sin\phi_f $      & - & - & - &$-2.9 \times 10^{-3}$ &$-1.6 \times 10^{-2}$ &$1.1 \times 10^{-2}$ \\
${f}\sin\delta_f $      & $0$ & $0$ & $0$ &$0$ &$0$ &$0$ \\
SM     & $0 $ & $0 $ & $0 $ &$0$ &$0$ &$0$ \\
\hline\hline
\end{tabular}
\caption{Integrated asymmetries for  $\tilde{d}$,  ${f}\sin\phi_f$ or ${f}\sin\delta_f$ $= 5 \times 10^{-4}~{\rm GeV}^{-1}$ and for the SM ($f, \tilde{d}=0$). }
\label{t:basic}
\end{table}
%%%%%%%%%%

We now consider two types of effects that can  dilute our estimated asymmetries. First, since the neutrinos are not detected, full reconstruction of the top four-momentum is not possible. As discussed above we deal with this complication by considering the asymmetries $\tilde{A}_i$ of Eq.~\ref{prodcoprime} instead of the $A_i$. Second, we want to introduce cuts to suppress known background and see their effect on the asymmetries. Specifically we know that the cross-section for $pp \to b\bar{b} \mu^+ \mu^- X$ (no intermediate $t\bar{t}$) within the SM is significantly larger. This process is dominated by strong production of $b\bar{b}$ pairs with the lepton pair attached to a photon or $Z$ boson. With the same default Madgraph parameters and cuts, we estimate this cross-section to be about 24 pb, or some 5.6 times larger than that for the process of interest to us. This SM process is $CP$ conserving so it cannot fake any of our $CP$ odd asymmetries. Nevertheless, it will make any non-zero $A_i$ smaller because it increases the total number of events. To reduce this background, we introduce two standard sets of cuts. The first is a set of acceptance and separation cuts for  the muons and $b$ quarks  are: 
\begin{eqnarray}
p_T(\mu^\pm) > 20~{\rm GeV}&& p_T(b,{\bar b}) > 25 ~{\rm GeV} 
\nonumber \\
|\eta(b,{\bar b},\mu^\pm)| < 2.5 && \Delta R (b{\bar b}) > 0.4.
\label{cuts1}
\end{eqnarray}
This set of cuts already reduces the background cross-section by a factor of 20, to about 1.2~pb. 
A second set of cuts includes Eq.~\ref{cuts1} and also requires missing transverse energy, $\slashed{E}_T$, to select events originating from semi-leptonic $t,\bar{t}$ decays,
\begin{eqnarray}
&&\slashed{E}_T > 30 ~{\rm GeV}.
\label{cuts2}
\end{eqnarray}
The requirement of a minimum $\slashed{E}_T$ completely eliminates the background in our parton level study, although this will no longer be the case when detector effects, initial state radiation and final state radiation are included. 

In Table~\ref{t:withcuts} we show the effect of these cuts in our asymmetries.
%%%%%%%%%%%%
\begin{table}[h]
\centering
\begin{tabular}{| c | c | c | c | c | c | c | c |}
\hline\hline
 & $A_1$ & $A_2$ & $A_3$ & $A_4$ & $A_5$ & $A_6$ & cuts \\
\hline 
$\tilde{d} $      & $0.1$ & $-2.2 \times 10^{-2}$ & $2.5 \times 10^{-3}$ &$-7.4 \times 10^{-2}$ &$4.1\times 10^{-2}$ &$-8.4 \times 10^{-3}$ & Eq.~\ref{cuts1} \\
    & $0.1$ & $-2.1 \times 10^{-2}$ & $2.9 \times 10^{-3}$ &$-7.5 \times 10^{-2}$ &$3.6\times 10^{-2}$ &$-6.4 \times 10^{-3}$ & Eqs.~\ref{cuts1},~\ref{cuts2} \\
\hline
${f}\sin\phi_f $      & - & - & - &$-5.3 \times 10^{-3}$ &$-1.6 \times 10^{-2}$ &$1.6 \times 10^{-2}$ & Eq.~\ref{cuts1}\\
     & - & - & - &$-5.8 \times 10^{-3}$ &$-1.7 \times 10^{-2}$ &$1.7 \times 10^{-2}$ & Eqs.~\ref{cuts1},~\ref{cuts2}\\
\hline\hline
\end{tabular}
\caption{Integrated asymmetries for  $\tilde{d}$ or ${f}\sin\phi_f$  $= 5 \times 10^{-4}~{\rm GeV}^{-1}$ with the cuts defined in Eqs.~\ref{cuts1},~\ref{cuts2}. }
\label{t:withcuts}
\end{table}
%%%%%%%%%%
Comparing Tables~\ref{t:basic}~and~\ref{t:withcuts} we see only small changes, with some asymmetries actually being larger after the cuts are imposed. This, of course, is due to a smaller denominator in the respective $A_i$ reflecting the smaller number of events. 
Within our approximations, the cross-section  becomes $2.6~{\rm pb}$ with the cuts of Eq.~\ref{cuts1}, and $2.3~{\rm pb}$ with both sets of cuts indicating a loss of $40-50\%$ in the total number of events.

\begin{figure}
\centering
\resizebox{!}{7cm}{\includegraphics{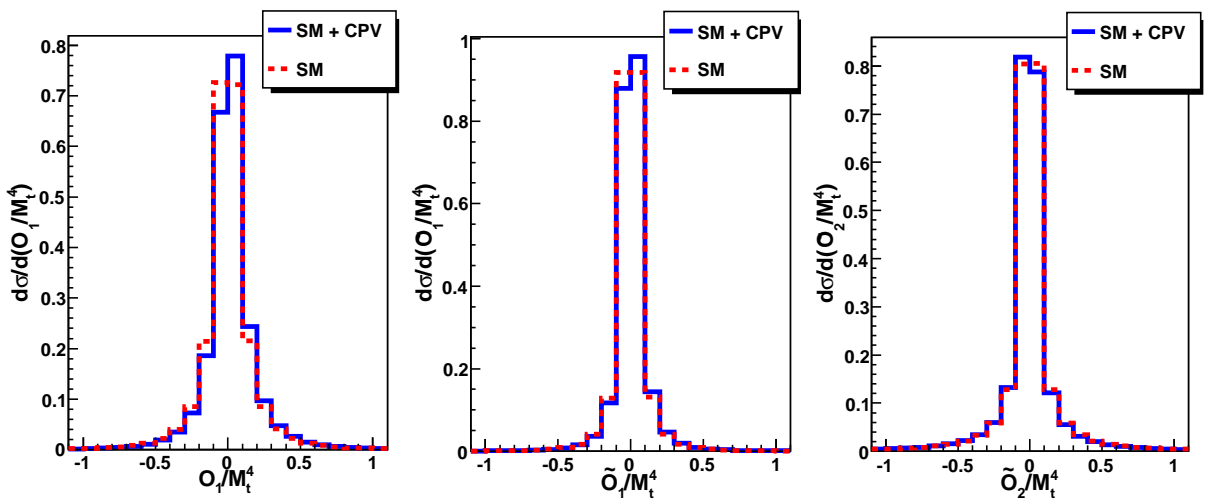}}
\caption{$d\sigma/d{\cal O}_1$ and $d\sigma/d{\tilde{\cal O}}_1$ distributions for the cases $\tilde{d}=0$ (SM) and $\tilde{d}=5\times 10^{-4}~{\rm GeV}^{-1}$ as well as $d\sigma/d{\tilde{\cal O}}_2$ for $f\sin\phi_f= 5 \times 10^{-4}~{\rm GeV}^{-1}$.}
\label{f:fig1}
\end{figure}

Next we calculate the asymmetries associated with the correlations $\tilde{\cal O}_{1,2}$ (which do not depend on full reconstruction of the top four-momentum) and show these results in Table~\ref{t:notfulltop}. For $\tilde{d}=5\times 10^{-4}$~GeV$^{-1}$ and ${f}\sin\phi_f=5\times 10^{-4}$~GeV$^{-1}$ we present results corresponding to the two sets of cuts.
%%%%%%%%%%%%
\begin{table}[h]
\centering
\begin{tabular}{| c | c | c | c | c |}
\hline\hline
 & $\tilde{A}_1$ & $\tilde{A}_2$ & $\tilde{A}_3$ & cuts \\
\hline 
$\tilde{d}$ & $5.6 \times 10^{-2}$ & $-4.1\times 10^{-3}$ & $1.8\times 10^{-2}$ & Eq.~\ref{cuts1} \\
& $5.5 \times 10^{-2}$ & $-3.5\times 10^{-3}$ & $1.8\times 10^{-2}$ & Eqs.~\ref{cuts1},~\ref{cuts2} \\
\hline
${f}\sin\phi_f$& $-5.4\times 10^{-3}$ & $-2.6\times 10^{-2}$ & $5.6\times 10^{-3}$ & Eq.~\ref{cuts2} \\
& $-6.2\times 10^{-3}$ & $-2.7\times 10^{-2}$ & $4.0\times 10^{-3}$ & Eqs.~\ref{cuts1},~\ref{cuts2} \\
\hline\hline
\end{tabular}
\caption{Integrated asymmetries without full top momentum reconstruction 
for  $\tilde{d}$ or ${f}\sin\phi_f$ $= 5 \times 10^{-4}~{\rm GeV}^{-1}$ with the cuts defined in Eqs.~\ref{cuts1},~\ref{cuts2}.  }
\label{t:notfulltop}
\end{table}
%%%%%%%%%%
The largest asymmetry for a $\tilde{d}$ coupling is $\tilde{A}_1$ which requires distinguishing between the $b$ and $\bar{b}$ jets. Interestingly the most promising asymmetry for probing $CP$ violation in the decay vertex is $\tilde{A}_2$, which does not require distinguishing the $b$ and $\bar{b}$ jets. To gain more insight into  these asymmetries we show the differential distributions $d\sigma/d{{\cal O}}_1$ and $d\sigma/d{\tilde{\cal O}}_1$ for $\tilde{d}= 5 \times 10^{-4}~{\rm GeV}^{-1}$ as well as $d\sigma/d{\tilde{\cal O}}_2$ for $f\sin\phi_f= 5 \times 10^{-4}~{\rm GeV}^{-1}$ in Figure~\ref{f:fig1}.
%--------------------------------------------------------------------
%--------------------------------------------------------------------

The distributions shown in Figure~\ref{f:fig1} can also be used to estimate the sensitivity of LHC to the anomalous couplings as was done for the case of $\tilde{d}$ and $\tilde{\cal O}_1$ in Ref.~\cite{Sjolin:2003ah}. 
In this paper we rely exclusively on the counting asymmetries since we are ignoring detector and other issues that are more important for the distributions. We summarize our results for the most promising asymmetries in terms of the dimensionless anomalous couplings
\begin{eqnarray}
d_t \equiv  \tilde{d} \, m_t, && 
f_t \equiv  f\, m_t
\end{eqnarray}
for $m_t = 171.2$~GeV as follows. The largest asymmetry is 
\begin{eqnarray}
A_1 & = & 1.17 \, d_t .
\label{largest}
\end{eqnarray}
Measurement of this asymmetry requires reconstruction of the top pair four-momenta. Assuming that only muon and $b$ four-momenta can be reconstructed, our asymmetries are
\begin{eqnarray}
\tilde{A}_1 &=& 0.64\, d_t -0.072\, f_t\sin\phi_f \nonumber \\
\tilde{A}_2 &=& -0.041\, d_t -0.32\, f_t\sin\phi_f \nonumber \\
\tilde{A}_3 &=& 0.21\, d_t +0.047\, f_t\sin\phi_f .
\label{notoprec}
\end{eqnarray}

We now estimate the $5\sigma$ sensitivity of LHC with an integrated luminosity of $10~fb^{-1}$ to the anomalous couplings. Using Eq.~\ref{staterror} with $N = 23k$  events per year (corresponding to $\sigma = 2.3$~pb) we see that $5\sigma$ sensitivity requires $A_i \geq 0.033$. From $\tilde{A}_1$ and $\tilde{A}_2$ respectively, we find setting only one anomalous coupling to be non-zero at a time, 
\begin{eqnarray}
|d_t| \geq  0.05,    && |\tilde{d}| \geq    3.0 \times 10^{-4}~{\rm GeV}^{-1}
  \\ \nonumber
|f_t \sin\phi_f| \geq   0.10,  && |f\sin\phi_f| \geq  6.0 \times 10^{-4}~{\rm GeV}^{-1}
\end{eqnarray}

The study of Ref.~\cite{Sjolin:2003ah} found that the LHC with an integrated luminosity of $10~fb^{-1}$ could achieve a $5\sigma$ sensitivity  to $d_t > 0.05$ by considering an observable proportional to $\tilde{A}_{1}$.  The agreement with our estimate is very good although the analysis in Ref.~\cite{Sjolin:2003ah} is much more complete. The conclusion by Sj\"olin in Ref.~\cite{Sjolin:2003ah} is reached by analyzing both the counting asymmetry and the distribution proportional to $d\sigma/d{\tilde {\cal O}}_1$ including both di-lepton and lepton plus jets channels from the $W^+$ and the $W^-$ in the semileptonic $t$ and $\bar{t}$ decays. His study also includes many background and reconstruction issues that we have completely ignored. However, we can roughly compare the two results: we have used only the channel  with $\mu^+$ and $\mu^-$ from the $W^+$ and $W^-$ decays and have looked only at the counting asymmetry. The comparable $5\sigma$ sensitivity result of Ref.~\cite{Sjolin:2003ah}, from the counting asymmetry in the di-lepton channel is based on 13k events that survive all the cuts. After our less restrictive cuts, our  $5\sigma$ sensitivity result is based on $23k$ events resulting in about $30\%$ more reach. We conclude that the two estimates are consistent with each other.

\section{Strong Interaction Phases}

In the previous sections we have constructed observables designed to isolate true $CP$ violation from the $T$ -odd triple products. However in some cases it may be desirable to  isolate $CP$ conserving but $T$-odd effects induced by strong interaction (unitarity) phases and we address this issue now. We begin by choosing two observables to construct counting asymmetries. From Table~\ref{t:notfulltop} it is clear that $\tilde{{\cal O}}_2$ is the most sensitive observable to phases in the decay vertex. We can easily construct a $CP$-even version of this observable by replacing the sum of $b$ and $\bar{b}$ jet momenta with their difference. Since  it is hard to distinguish these two jets, we would like to have at least one example of a $CP$ even observable that does not require this separation. Thus we consider
\begin{eqnarray}
{\cal O}_a &=& \, \tilde{q}\cdot (p_{\mu^+}+p_{\mu^-}) \,\epsilon(p_{\mu^+},p_{\mu^-},p_b+p_{\bar{b}},\tilde{q}) \nonumber \\
{\cal{O}}_b &=& \, \tilde{q}\cdot (p_{\mu^+}-p_{\mu^-}) \,\epsilon(p_{\mu^+},p_{\mu^-},p_b-p_{\bar{b}},\tilde{q}).
\label{cpeven}
\end{eqnarray}
Proceeding as in the previous section we find results shown in Table \ref{t:stph}.
%%%%%%%%%%%%
\begin{table}[h]
\centering
\begin{tabular}{| c | c | c |}
\hline\hline
${A}_a$ & ${A}_b$  & cuts \\
\hline 
$4.2 \times 10^{-3}$ & $-3.1 \times 10^{-2}$ & Eq.~\ref{cuts1} \\
$3.0 \times 10^{-3}$ & $-2.7 \times 10^{-2}$ & Eqs.~\ref{cuts1},~\ref{cuts2} \\
\hline\hline
\end{tabular}
\caption{Integrated $CP$-even asymmetries with $f\sin\delta_f = 5 \times 10^{-4}~{\rm GeV}^{-1}$  with the cuts defined in Eqs.~\ref{cuts1},~\ref{cuts2}.  }
\label{t:stph}
\end{table}
%%%%%%%%%%

Using the larger of these two asymmetries we can write
\begin{eqnarray}
A_b &=& -0.32 f_t \sin\delta_f,
\label{cpeveneq}
\end{eqnarray}
from which we conclude that the LHC with $10~{\rm fb}^{-1}$ will have a $5\sigma$ sensitivity to  
\begin{eqnarray}
|f_t\sin\delta_f| \geq  0.10  && |f\sin\delta_f| \geq 6.0 \times 10^{-4}~{\rm GeV}^{-1}.
\end{eqnarray}

The magnetic transition factor $f$ is induced by one-loop QCD corrections at the level $f = 8\times 10^{-5}~{\rm GeV}^{-1}$ \cite{Li:1990qf}, and it has been claimed that this level can be reached at LHC \cite{delAguila:2002nf}. The $1\sigma$ contraint for 10~fb$^{-1}$ has been estimated at $-2.6 \times 10^{-4} ~{\rm GeV}^{-1} \leq f \leq 1.4 \times 10^{-4}~{\rm GeV}^{-1}$ by setting other anomalous couplings to zero in 
Ref.~\cite{AguilarSaavedra:2007rs}. This constraint was obtained using $T$-even observables and is somewhat better than what we find with the $T$-odd correlation ${\cal O}_a$. Of course, our $T$-odd observables require a non-zero absorptive phase $\delta_f$, unlike the observables used in the study of Ref.~\cite{AguilarSaavedra:2007rs}.  Although it is easy to see that QCD corrections to the tree-level weak vertex can introduce such a phase, it has not been calculated yet to our knowledge.

\section{Summary and Conclusion}

We have presented the results of a numerical implementation of the results in Ref.~\cite{Antipin:2008zx} using MADGRAPH for event generation at the parton level. We have used these events to estimate the sensitivity of the LHC to $CP$ violating anomalous top-quark couplings.

For the case of  the coupling $d_t$ that parametrizes $CP$ violation in the $t\bar{t}$ production process, we find that the LHC with 10~fb$^{-1}$ can rule out values larger than $0.05$ at the $5\sigma$ level. We reach this conclusion by considering only counting asymmetries in the di-muon channel. Specifically, our result follows from $\tilde{A}_1$ which is the most sensitive asymmetry that does not require complete reconstruction of the $t$ and $\bar{t}$ four-momenta. Our result is consistent with the estimates of Sj\"olin in Ref.~\cite{Sjolin:2003ah}. Further refinements to our study that can be carried out by the experimental collaborations along the lines of  Ref.~\cite{Sjolin:2003ah} include: going beyond the parton level to include detector effects; using lepton+jets channels; using distribution shapes in addition to the counting asymmetries. Our paper has shown that there are other asymmetries, beyond the one studied in Ref.~\cite{Sjolin:2003ah}, that can provide comparable levels of sensitivity, and/or different handles on the analysis. For example the correlation $\tilde{A}_1$ used by Sj\"olin requires distinguishing between the $b$ and $\bar{b}$ jets. Although this may be possible, as discussed in Ref.~\cite{Sjolin:2003ah}, we have presented an alternative observable, $\tilde{A}_2$, where this is not necessary. 

For the case of the anomalous coupling $f_t\sin\phi_f$ that parametrizes $CP$ violation in the decay vertex, we find that the LHC with 10~fb$^{-1}$ can rule out values larger than $0.10$ at the $5\sigma$ level. This result is based on the counting asymmetry $\tilde{A}_2$. Using Ref.~\cite{Sjolin:2003ah} as a guide, we may expect a loss of a factor 2-3 in sensitivity when going from our parton level study to a detector level one.

For the case of $CP$ even but $T$ odd correlations induced by strong phases in top decay, we find that the LHC with 10~fb$^{-1}$ can test values as small as $f_t\sin\delta_f = 0.10$ at the $5\sigma$ level. This estimate is consistent with that of Ref.~\cite{AguilarSaavedra:2007rs} for $f_t$ using different observables.

We have shown that, at least at the parton level, it is possible to significantly reduce or even eliminate the dominant background with a simple set of cuts that have a relatively small effect on the signal. $CP$ conserving background can arise from the fact that the initial state  at LHC ($pp$) is not a $CP$ eigenstate. This issue has not been addressed in detail so far, but a simple estimate in Ref.~\cite{Atwood:1992vj} puts it several orders of magnitude below the sensitivity that we have found.

\begin{acknowledgments}

This work was supported in part by DOE under contract number DE-FG02-01ER41155. We thank Fabio Maltoni for his guidance with  MADGRAPH. We are grateful to David Atwood, James Cochran and Soeren Prell  for useful conversations.

\end{acknowledgments}

\appendix

\section{Form Factors}

The form factors for CP violation in the production vertex appear in Ref.~\cite{Antipin:2008zx} split into three different contributions. The three contributions correspond to those of different $gg\to t\bar{t}$ diagrams and were split that way to facilitate checking them. In this paper we are only interested in the total result for use in numerical routines. This corresponds to Eq.~16 of Ref.~\cite{Antipin:2008zx} which we write here explicitly:
\begin{eqnarray}
&&C_1= \frac{\tilde{d}K_{\ell \ell} m_t}{6s^2(s^2-(t-u)^2)^2}
 \left(9(t-u)^6-2s^2(t-u)^4-7s^4(t-u)^2 \right.
 \\ &&\left. +32m_t^4(7s^4+9s^2(t-u)^2)+4m_t^2(7s^5+7(t-u)^2s^3+18(t-u)^4s)\right) \nonumber \\
&&C_3 = \frac{\tilde{d}K_{\ell \ell} m_t}{6s^2(s^2-(t-u)^2)^2}
\left(-7s^4-2(t-u)^2s^2+9(t-u)^4+4m_t^2(23s^3+9(t-u)^2s)\right)\nonumber
\label{prodff}
\end{eqnarray}
and $C_2= \frac{s}{2}\, C_3$. Notice that these form factors differ form those defined in Ref.~\cite{Antipin:2008zx} by factors of $t-u$ according to Eq.~\ref{prodco}.

The $Q_t$ and $Q_{\bar t}$ that appear in Eq.~\ref{asymcpdec} are linear combinations of available momenta that act as spin analyzers for the $t$ and $\bar{t}$ respectively. They were also given in  Ref.~\cite{Antipin:2008zx} separately for three different contributions from the different diagrams in $gg\to t\bar{t}$. Once again, we are only interested in the sum of these contributions in this paper so we present  it here explicitly for convenience:
\begin{eqnarray}
  Q_{t} &=& K_{\ell \ell} \, \frac{m_t}{ 6s^2(s^2-(t-u)^2)^2} \left\{\left(
  -32(7s^4+9(t-u)^2s^2)m_t^4+8s(s^2-(t-u)^2)
  \right . \right. \nonumber \\
  &&  \left. \cdot (7s^2+27(t-u)^2)m_t^2 +(29s^2-45(t-u)^2)(s^2-(t-u)^2)^2 \right) p_{\ell^-}
 \nonumber \\
  && + 2 \left( -s(29s^4-74(t-u)^2s^2+45(t-u)^4+4m_t^2(7s^3+9(t-u)^2s)) p_{\ell^-}\cdot p_t \right. \nonumber \\
  &&+ s(9(t-u)^4+2s(54m_t^2+17s)(t-u)^2+(84m_t^2-43s)s^3)p_{\ell^-}\cdot p_{\bar t} \nonumber \\
&&+\left.  (t-u)(45(t-u)^4+2(54m_t^2-37s)s(t-u)^2+s^3(29s-44m_t^2))p_{\ell^-}\cdot q \right) p_{\bar t} \nonumber \\
&&+  2\left( (t-u)(-45(t-u)^4-2(54m_t^2-37s)s(t-u)^2+(44m_t^2-29s)s^3)p_{\ell^-}\cdot p_{t} \right. \nonumber \\
&&+(t-u)(-45(t-u)^4-2(54m_t^2-55s)s(t-u)^2+(44m_t^2-65s)s^3)p_{\ell^-}\cdot p_{\bar t} \nonumber \\
&&+ \left.\left. s(45(t-u)^4+2(18m_t^2-37s)s(t-u)^2+s^3(28m_t^2+29s))p_{\ell^-}\cdot q\right)q \right \},
\label{cpdecsa}
\end{eqnarray}
\begin{eqnarray}
 Q_{\bar t} &=& K_{\ell \ell} \, \frac{m_t}{ 6s^2(s^2-(t-u)^2)^2} \left\{\left(
  -32(7s^4+9(t-u)^2s^2)m_t^4+8s(s^2-(t-u)^2)
  \right . \right. \nonumber \\
  &&  \left. \cdot (7s^2+27(t-u)^2)m_t^2 +(29s^2-45(t-u)^2)(s^2-(t-u)^2)^2 \right) p_{\ell^+}
 \nonumber \\
  && + 2 \left( -s(29s^4-74(t-u)^2s^2+45(t-u)^4+4m_t^2(7s^3+9(t-u)^2s)) p_{\ell^+}\cdot p_{\bar t} \right. \nonumber \\
  &&+ s(9(t-u)^4+2s(54m_t^2+17s)(t-u)^2+(84m_t^2-43s)s^3)p_{\ell^+}\cdot p_{t} \nonumber \\
&&-\left.  (t-u)(45(t-u)^4+2(54m_t^2-37s)s(t-u)^2+s^3(29s-44m_t^2))p_{\ell^+}\cdot q \right) p_{t} \nonumber \\
&&+  2\left(- (t-u)(-45(t-u)^4-2(54m_t^2-37s)s(t-u)^2+(44m_t^2-29s)s^3)p_{\ell^+}\cdot p_{\bar t} \right. \nonumber \\
&&-(t-u)(-45(t-u)^4-2(54m_t^2-55s)s(t-u)^2+(44m_t^2-65s)s^3)p_{\ell^+}\cdot p_{ t} \nonumber \\
&&+ \left.\left. s(45(t-u)^4+2(18m_t^2-37s)s(t-u)^2+s^3(28m_t^2+29s))p_{\ell^+}\cdot q \right) q\right \}.
\label{cpdecs}
\end{eqnarray}
The factor $K_{\ell\ell}$ defined in Ref.~\cite{Antipin:2008zx} is given by:
\begin{eqnarray}
K_{\ell\ell} &\equiv & 16\, (\pi^2\alpha_s^2g^8 )\, \left(p_b\cdot p_\nu\right)\left( p_{\bar{b}}\cdot p_{\bar{\nu}} \right)\, \left(\frac{\pi}{m_t\Gamma_t}\right)^2\left(\frac{\pi}{M_W\Gamma_W}\right)^2 \nonumber \\
&\times&  \delta(p_t^2-m_t^2)\delta(p_{\bar{t}}^2-m_t^2) 
\delta(p_{W^+}^2-M_W^2)  \delta(p_{W^-}^2-M_W^2).
\end{eqnarray}
For our numerical implementation we rewrite all four delta functions as the respective Breit-Wigner distributions behind them, for example:
\begin{eqnarray}
 \left(\frac{\pi}{m_t\Gamma_t}\right) \delta(p_t^2-m_t^2) \to 
 \frac{1}{(p_t^2-m_t^2)^2+\Gamma_t^2m_t^2}.
\end{eqnarray}

\section{Small Asymmetries and Statistical Fluctuations}
\label{a:stat}

For our numerical analysis of the counting asymmetries in Eq.~\ref{asym} it is important to know the level at which these can be faked by statistical fluctuations. For a $CP$ conserving background (such as the SM) which does not generate the asymmetry, the probability for a given event to fall in the bin with ${\cal O}_i>0$ is $p=1/2$. The number of events falling in this bin out of a total number of events $N$, follows a binomial distribution with mean $Np$ and standard deviation $\sqrt{Np(1-p)}$. It follows that $N$ events due to a $CP$ conserving process lead to the expectation
\begin{eqnarray}
{A_i} = 0 \pm \left(\frac{1}{\sqrt{N}}\right)_{stat}.
\label{staterror}
\end{eqnarray}

Several asymmetries were found to be zero in this paper and we now show that those asymmetries are really zero and not just numerically small. To do this we show results obtained by varying the number of generated events and/or the size of the anomalous couplings to demonstrate that certain small asymmetries are consistent with expectations from statistical fluctuations, and not with $CP$ violation.  
For the SM case we show in Table~\ref{t:SM} the asymmetries that we obtain when generating $10^6$ and $10^7$ events with MADGRAPH using the cuts in Eqs.~\ref{cuts1},~\ref{cuts2}. For comparison we show the $3\sigma$ statistical error from Eq.~\ref{staterror} above. 
%%%%%%%%%%%
\begin{table}[h]
\centering
\begin{tabular}{| c | c || c | c | c | c | c | c | c | c | c |}
\hline\hline
N & $\frac{3}{\sqrt{N}}$ &$ A_1$ & $ A_2$ & $ A_3$ & $ A_4$ & $ A_5$ & $ A_6$ & $ \tilde{A}_1$ & $ \tilde{A}_2$ & $ \tilde{A}_3$\\
\hline 
$10^6$ &$3.0$   & $-1.6$ & $0.9$ & $1.1$ & $-0.8$ & $-2.1$ & $3.6$ & $0.5$ & $-0.5$ & $0.9$ \\
$10^7$ & $0.95$    & $-0.1$ & $0.0$ & $0.2$ & $-0.4$ & $-0.1$ & $0.0$ & $0.1$ & $-0.1$ & $0.6$ \\
\hline\hline
\end{tabular}
\caption{Integrated asymmetries with $10^6$ and $10^7$ generated SM events in units of $10^{-3}$. For comparison we show in the second column the expected $3\sigma$ statistical error also in units of $10^{-3}$.}
\label{t:SM}
\end{table}
%%%%%%%%%%
The results in Table~\ref{t:SM} show that at the $99\%$ confidence level the SM does not induce any $CP$-odd asymmetries.

We next turn our attention to the case of strong phases. We show in Table~\ref{t:strph} the asymmetries obtained by generating $10^6$ events with Madgraph using the cuts in Eqs.~\ref{cuts1},~\ref{cuts2}, for two different values of $f\sin\delta_f$.
%%%%%%%%%%%
\begin{table}[h]
\centering
\begin{tabular}{| c | c || c | c | c | c | c | c | c | c | c |}
\hline\hline
$f\sin\delta_f$ & $\frac{3}{\sqrt{N}}$ &$ A_1$ & $ A_2$ & $ A_3$ & $ A_4$ & $ A_5$ & $ A_6$ & $ \tilde{A}_1$ & $ \tilde{A}_2$ & $ \tilde{A}_3$\\
\hline 
$5\times 10^{-4}~{\rm GeV}^{-1}$ &$3.0$   & $0.3$ & $-2.1$ & $1.4$ & $-0.6$ & $-0.2$ & $0.3$ & $1.3$ & $2.0$ & $0.9$ \\
$5\times 10^{-3}~{\rm GeV}^{-1}$ & $3.0$    & $1.3$ & $-1.0$ & $-0.2$ & $-1.6$ & $2.2$ & $0.1$ & $1.2$ & $-1.3$ & $1.3$ \\
\hline\hline
\end{tabular}
\caption{Integrated asymmetries with $10^6$  generated events in units of $10^{-3}$ for two different values of $f\sin\delta_f$. For comparison we show in the second column the expected $3\sigma$ statistical error also in units of $10^{-3}$.}
\label{t:strph}
\end{table}
%%%%%%%%%%
If any of the numbers obtained with $f\sin\delta_f=5\times10^{-4}~{\rm GeV}^{-1}$ arose from this coupling and not from statistical fluctuations, we would see its value increase by a factor of ten for $f\sin\delta_f=5\times10^{-3}~{\rm GeV}^{-1}$ as the asymmetries are linear in the anomalous couplings. Table~\ref{t:strph} thus shows, as expected, that strong scattering phases cannot fake the truly $CP$-odd observables we have defined. 

Finally we consider the case of the asymmetries $A_{1,2,3}$ as related to $CP$ violation in the decay vertex. We show in Table~\ref{t:cpdec123} the results of simulations with $10^7$ events using the cuts in Eqs.~\ref{cuts1},~\ref{cuts2}, for two different values of $f\sin\phi_f$.
%%%%%%%%%%%
\begin{table}[h]
\centering
\begin{tabular}{| c | c || c | c | c | }
\hline\hline
$f\sin\phi_f$ & $\frac{3}{\sqrt{N}}$ &$ A_1$ & $ A_2$ & $ A_3$ \\
\hline 
$5\times 10^{-4}~{\rm GeV}^{-1}$ &$0.95$   & $2.5$ & $-0.4$ & $-1.7$ \\
$5\times 10^{-3}~{\rm GeV}^{-1}$ & $0.95$    & $23$ & $-5.1$ & $-10.1$ \\
\hline\hline
\end{tabular}
\caption{Integrated asymmetries with $10^6$  generated events in units of $10^{-3}$ for two different values of $f\sin\phi_f$. For comparison we show in the second column the expected $3\sigma$ statistical error also in units of $10^{-3}$.}
\label{t:cpdec123}
\end{table}
%%%%%%%%%%
These results appear to show that these asymmetries are not zero but rather very small. The expected linear scaling with the size of the anomalous coupling is not satisfied perfectly leading us to think that statistical fluctuations are not completely under control in this case.
As stated in the main text, however, there are better asymmetries to measure this coupling so we do not pursue the issue further.

To summarize, we have  performed the following checks on our numerical results:
\begin{itemize}

\item By generating event samples of different sizes ranging from $10^5$ to $10^7$ for different values of $\tilde{d},$ and $f$, we have verified that the asymmetries scale linearly with the respective anomalous coupling  and that the statistical error scales as the square root of the number of events. 
\item Using the form factors corresponding to $CP$ violation due to a Higgs boson we have reproduced the asymmetry in Ref.~\cite{Valencia:2005cx}.
\item For the case of $\tilde{d}$ and asymmetry $\tilde{A}_1$ we have verified that we are in substantial agreement with the much more detailed analysis of Ref.~\cite{Sjolin:2003ah}.
\item For the case of strong interaction phases induced by QCD corrections we have checked that we are in rough agreement with the estimates in Ref.~\cite{AguilarSaavedra:2007rs} which use different observables.

\end{itemize}

\end{document}